\documentclass{article}
\usepackage{spconf,amsmath,graphicx}
\usepackage{multirow}

\usepackage{tikz}
\usetikzlibrary{positioning}

\title{IMPROVING SPEECH RECOGNITION ACCURACY OF LOCAL POI USING GEOGRAPHICAL MODELS}
\name{BLIND}
\address{BLIND}

\name{Songjun Cao$^*$, Yike Zhang$^*$, Xiaobing Feng, Long Ma \thanks{ $^*$ Both authors contributed equally to this work.}}
\address{Tencent Cloud Xiaowei, Beijing, China}

\begin{document}
%
\maketitle
\begin{abstract}
Nowadays voice search for points of interest (POI) is becoming increasingly popular. 
However, 
speech recognition for local POI names still remains a challenge due to multi-dialect and long-tailed distribution of POI names. 
This paper improves speech recognition accuracy for local POI from two aspects.
Firstly, 
a geographic acoustic model (Geo-AM) is proposed. 
The proposed Geo-AM deals with multi-dialect problem using dialect-specific input feature and dialect-specific top layers.
Secondly, 
a group of geo-specific language models (Geo-LMs) are integrated into our speech recognition system 
to improve recognition accuracy of long-tailed and homophone POI names.
During decoding,
a specific Geo-LM is selected on-demand according to the user's geographic location. 
Experiments show that 
the proposed Geo-AM achieves 6.5\%$\sim$10.1\% relative character error rate (CER) reduction on an accent test set and
the proposed Geo-AM and Geo-LMs totally achieve over 18.7\% relative CER reduction on a voice search task for Tencent Map.

\end{abstract}
%


\section{Introduction}
\label{sec:intro}
Automatic speech recognition (ASR) provides a more natural way to human-machine interaction (HMI).
The point of interest (POI) search with voice is one of typical HMI scenes. 
Assuming you are driving a car on the road and don't know how to reach your destination,
you can give a voice command to a map app to set your destination and start navigation.
However,
although deep learning techniques improve speech recognition accuracy by a large margin recently,
there are still some challenging problems for local POI recognition.

This paper mainly focuses on Chinese POI recognition.
In China,
dialects vary from geographical regions to geographical regions. 
Although different dialects may share some similarities, 
there are obvious differences at the phonological level. 
As a result, 
ASR system trained on many dialects simultaneously may fail to generalize well for each individual dialect.

There are massive POI names in China. 
The total amount of POI names to be searched in a navigation system is usually more than 1,200 million. 
Since POI names follow a long-tailed distribution, 
it is ineffective to model infrequent POI names using a general language model (LM).
Another difficulty for POI recognition is that 
there are many homophone POI names, which is especially serious in China. 

In order to alleviate the above problems in POI recognition, 
geographical location information is used for both acoustic modeling and language modeling in this paper.
The main contributions of this paper are as follows:
\begin{itemize}
\item
This paper proposes a geographical acoustic model (Geo-AM) to deal with the multi-dialect problem.
Dialects are usually specific to geographical regions or social groups. 
Therefore, the Geo-AM encodes users' geographical location into a dialect-specific vector as an additional input feature. 
Furthermore, 
the Geo-AM introduces multiple dialect-specific top layers, 
each of which corresponds to a dialect region. 
With dialect-specific top layers,
the proposed Geo-AM can efficiently exploit geographical information while keeping flexibility for further optimization.

\item
Generally, users are only interested in nearby POI names. 
In a specific region, 
the total number of POI names is much smaller and there are less homophone POI names. 
Starting from this, 
we build and integrate a group of geo-specific language models (Geo-LMs) into the ASR system 
to improve the recognition accuracy of long-tailed and homophone POI names.
During decoding, 
a specific Geo-LM will be selected on-demand according to geographic location information, 
which is attached to user queries. 
To further improve recognition accuracy, 
the n-best rescoring is done with a neural network LM combined with another group of Geo-LMs built for rescoring.
\end{itemize}

The rest of this paper is organized as follows. 
In Section~\ref{sec:related_works}, we discuss some related works on multi-dialect speech recognition and POI recognition. 
Section~\ref{sec:baseline} describes the details of the baseline ASR system used in this paper. 
Geo-AM and Geo-LMs are described in Section~\ref{sec:am} and Section~\ref{sec:lm} respectively. 
Section~\ref{sec:exp} shows the experimental results and analysis. Finally, 
Section~\ref{sec:cons} concludes this paper.


\section{Related Works}
\label{sec:related_works}
Recently, 
there have been some attempts to solve the multi-dialect problem in speech recognition, 
which mainly fall into two ways: ``multi-model'' and ``single-model''.
In multi-model approaches,
an individual AM is trained for each dialect when enough data is available for each dialect. 
When dialect-specific data is scarce, 
Huang {\it et al.} \cite{huang2014multi} and Chen {\it et al.} \cite{chen2015improving} provide a solution that 
jointly train an universal AM which will be fine-tuned with dialect-specific data to get dialect-specific AMs. 
In single-model approaches,
a single AM is trained to deal with all dialects. 
Some of them feed dialect-related features, 
such as I-vectors \cite{dehak2010front} or dialect information \cite{li2018multi, jain2018improved, yoo2019highly},
into AMs to deal with the dialect problem. 
Some researchers introduce multi-task learning into multi-dialect speech recognition. Yang {\it et al.} 
\cite{yang2018joint} adopts dialects classification as the secondary learning task. 
Compared with multi-model approaches, single-model approaches are usually more efficient but less flexible.

In this paper, 
we first construct a ``single-model'' baseline like \cite{li2018multi}.
In order to further optimize the corresponding component for a specific dialect when additional data is available,
dialect-specific top-layers are further introduced into the proposed model.

An efficient way to improve speech recognition accuracy of POI names is to 
utilize geo-location dependent LMs \cite{heerden2009language, stent2009geo, chelba2015geo, xiao2018geographic}. 
For each user,
Sten {\it et al.} \cite{stent2009geo} trains a Geo-LM dynamically using nearby POI names and 
combines the Geo-LM with a baseline LM before or at decoding. 
In \cite{xiao2018geographic}, 
a class-based Geo-LM is constructed dynamically for each user depending on users' geographic location,
within a difference-LM based weighted finite state transducer (WFST) system.
All above approaches construct LMs or WFSTs on-the-fly according to users' geographical locations, 
which is time consuming and hard to incorporate plenty of POI names into a Geo-LM. 
Moreover, 
the class-based Geo-LMs can only deal with pre-defined grammars.

In this paper, 
for each pre-defined region a Geo-LM is pre-trained and Geo-LMs are dynamically combined with a baseline LM during decoding. 
In addition, 
prior works mainly integrate Geo-LMs into first-pass decoding. 
To further improve recognition accuracy, 
we integrate Geo-LMs into both first-pass decoding and n-best re-scoring. 


\section{Baseline System}
\label{sec:baseline}
The baseline AM is trained with lattice-free maximum mutual information (LF-MMI) \cite{povey2016purely} criterion. It consists of two CNN layers and three TDNN-OPGRU \cite{cheng2018output} blocks, which interleave TDNN and output-gate PGRU (OPGRU) layers. 
Besides, SpecAugment \cite{park2019specaugment} algorithm is used to improve the robustness of the AM. 

The baseline LM is a word-level Kneser–Ney smoothed 5-gram model. For further improvement, a character-level Kneser–Ney smoothed 5-gram LM and a QRNN \cite{bradbury2016quasi} model are used to rescore n-best lists of the first-pass decoding output. 


\section{Geographical Acoustic Model}
\label{sec:am}
\subsection{Dialect-specific Input Feature}
\label{ssec:am_inp}
From the aspect of linguistics, China can be divided into several dialect regions. 
People from some adjacent provinces usually have similar acoustic characteristics.
Therefore,
such provinces can be clustered into one dialect region. 

\begin{figure}[htb]
  \centerline{\includegraphics[width=0.5\textwidth]{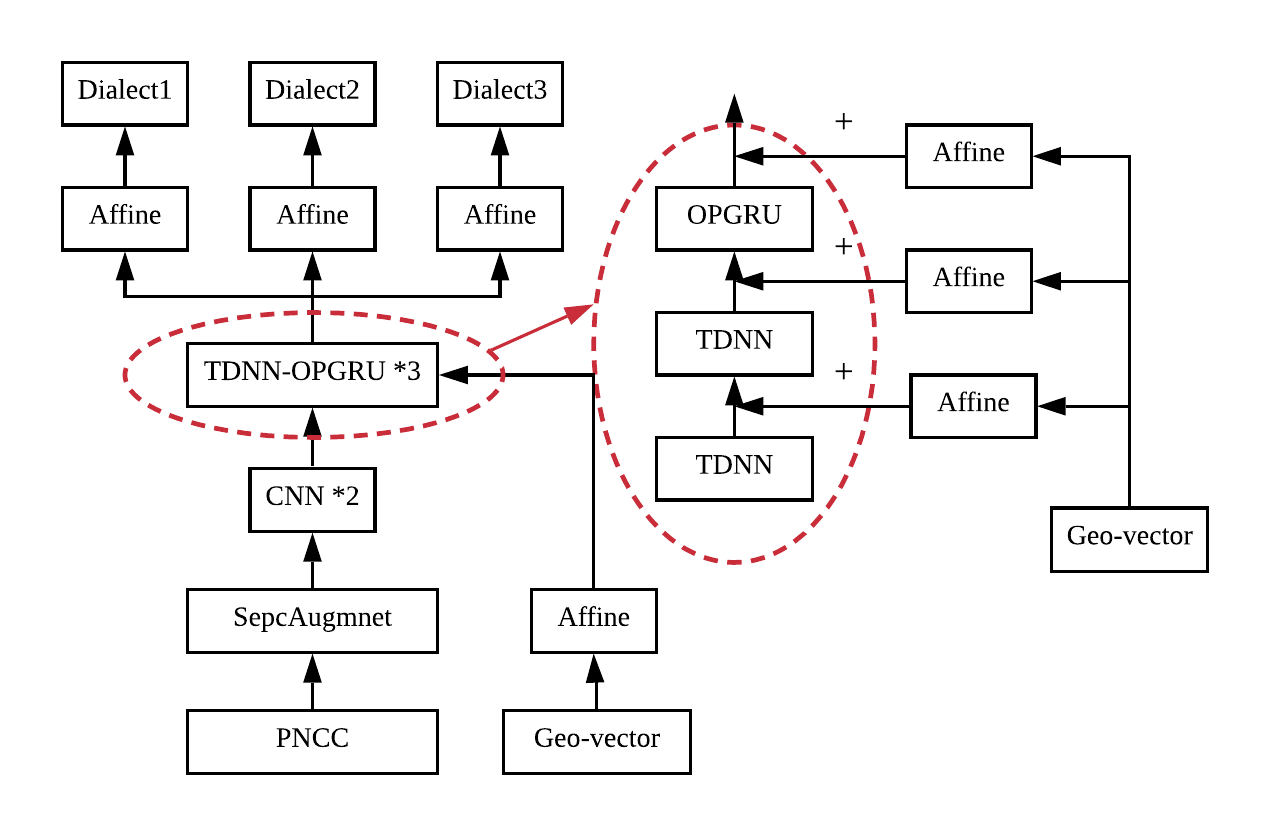}}
\caption{Architecture of the proposed Geo-AM.}
\label{fig:geo_am}
\end{figure}

Like \cite{li2018multi}, 
we represent dialect information as a one-hot vector (Geo-vector)
which will be served as an additional feature fed into the AM as depicted in Figure \ref{fig:geo_am}. 
Dialect-specific input feature will be transformed by an affine layer before added to the output of TDNN or OPGRU layers. 
In this way, 
the proposed Geo-AM can utilize additional dialect information at both training and inference stage.

\subsection{Dialect-specific Top Layer}
\label{ssec:am_op}
The model mentioned in Subsection \ref{ssec:am_inp} can achieve gains generally over the baseline model 
due to the additional supervised information. 
However, 
it is hard to improve the accuracy of a specific dialect while 
maintaining the performance on other dialects. 
This may be attributed to that
dialect-specific information and dialect-independent information are coupled together in that model.
In order to make the Geo-AM more flexible, 
here we introduce dialect-specific layers into the Geo-AM.

As found in \cite{huang2014multi, chen2015improving}, 
the top layer can capture dialect information. 
Here we introduce dialect-specific top layers into the proposed Geo-AM as depicted in Figure \ref{fig:geo_am}. 
Each dialect has its own top layer which is adapted from the model in Subsection \ref{ssec:am_inp}. 
During the adaptation training, 
only top layer's parameters update while other parameters are fixed, 
which is easy for deployment.


\section{Geographical Language Model}
\label{sec:lm}
For simplicity,
this paper divides China into 34 local regions at province-level.
For each region, 
a word-level Geo-LM and a character-level Geo-LM are trained with local POI names in that region.

\subsection{Geo-LMs in first-pass decoding}
\label{ssec:lm1}
The underlying ASR system is based on a WFST based decoder,
which employs the difference LM principle as follows:
\begin{equation}
   H C L G_{bi} \circ F
   \label{eq:hclg}
\end{equation}
where $\circ$ denotes on-the-fly composition,
$H$ contains HMM definitions,
$C$ represents the context dependency, 
$L$ is the lexicon, 
$G_{bi}$ 
is a small LM consisting only of uni-grams and bi-grams of the baseline 5-gram LM,
and
\begin{equation}
   F = G_{bi}^{-} \circ G_{b} \circ G_{l}
   \label{eq:otf}
\end{equation}
where $G_{bi}^{-}$ is negated score version of $G_{bi}$,
$G_{b}$ is the 5-gram baseline LM and
$G_{l}$ is a Geo-LM.

For each query,
we first get the province, in which the user locates,
by location based services (LBS).
Then the corresponding Geo-LM is selected to do on-the-fly composition according to Eq.(\ref{eq:hclg}) and Eq.(\ref{eq:otf}).
As a result,
the probability of a word $w$ in first-pass decoding is 
\begin{equation}
    P^{(1)}(w|h) = \lambda P_{b}^{(1)}(w|h) + (1 - \lambda) P_{l}^{(1)}(w|h) 
    \label{eq:lm1}
\end{equation}
where $h$ is context,
$P_{b}^{(1)}(w|h)$ is the probability from the baseline LM,
$P_{l}^{(1)}(w|h)$ is the probability from a Geo-LM and 
$\lambda$ is a scalar that controls the contribution of different LMs.

\subsection{Geo-LMs in n-best rescoring}
\label{ssec:lm2}
In order to further improve recognition accuracy of local POI names,
a QRNN \cite{bradbury2016quasi} model is used to rescore n-best lists of the first-pass decoding outputs.
A single neural model usually fails to model long-tailed POI names.
Moreover,
it is impractical to train a geographical neural model for each region 
due to data sparsity.
Therefore,
we also incorporate a group of character-level Geo-LMs into the process of n-best rescoring.
Like Eq.(\ref{eq:lm1}), 
the probability of the word $w$ in second-pass decoding is 
\begin{equation}
\begin{aligned}
    P^{(2)}(w|h) &= \alpha P_{b}^{(2)}(w|h) + \beta P_{r}^{(2)}(w|h) \\
        & \qquad \qquad + (1 - \alpha -\beta) P_{l}^{(2)}(w|h) 
    \label{eq:lm2}
\end{aligned}
\end{equation}
where $P_{b}^{(2)}(w|h)$ is the probability from the character-level baseline LM,
$P_{r}^{(2)}(w|h)$ is the probability from the QRNN model,
$P_{l}^{(2)}(w|h)$ is the probability from a character-level Geo-LM, 
$\alpha$ and $\beta$ control the contribution of different LMs.

After getting probabilities $ P^{(1)}(w|h)$ and $ P^{(2)}(w|h)$,
the final probability of the word $w$ is
\begin{equation}
    P(w|h) = \gamma P^{(1)}(w|h) + (1 - \gamma) P^{(2)}(w|h) 
    \label{eq:lm}
\end{equation}
where $\gamma$ is a constant.


\section{Experiments}
\label{sec:exp}
We train AMs on hand-transcribed,
anonymized utterances from our production including Tencent Map. 
Our training data are collected from all regions of China, 
which is amounted to about 20K hours. 
Only one fifth training data have region information.
In all experiments,
40-dimensional PNCC \cite{kim2016power} is
 used as acoustic feature.

Both the word-level baseline LM and the character-level baseline LM are trained with 1,200M POI names collected from Tencent map.
In order to limit the model size, 
the baseline LMs are trained with large cutoffs 0-3-5-10-15. 
As a result,
many long-tailed or infrequent POI names are excluded from the baseline LMs.
For each province,
a word-level Geo-LM and a character-level Geo-LM are trained with local POI names in that province 
collected from Tencent map and the Internet.
The amount of training data for Geo-LMs varies from 30K POI names to 12.6M POI names.
The Geo-LMs are trained with small (standard) cutoffs 0-2-2-2-2, 
which keeps more long-tailed POI names in Geo-LMs.
In addition,
the QRNN model adopts a adaptive softmax output layer \cite{joulin2017efficient} to reduce computational complexity.

Based on the dialect regions and users' distribution,
this paper splits China into 10 dialect regions as shown in Table~\ref{tab:region}. 
Table~\ref{tab:region} also gives the amount of training corpus for each region.
\begin{table}[htp]
\caption{Dialect region division and the amount of training corpus (in hours) for each region.}
\centering
\begin{tabular}{clc}
\hline
\textbf{Regions} & \textbf{Provinces}  & \textbf{Corpus} \\
\hline
1  &  Zhejiang Jiangsu            & 522  \\ 
2  &  Sichuan Chongqing Guizhou   & 345  \\ 
3  &  Shandong Henan              & 598  \\ 
4  &  Heilongjiang Jilin Liaoning & 372  \\ 
5  &  Guangdong                   & 450  \\ 
6  &  Shanxi Gansu Shaanxi        & 247  \\ 
7  &  Hunan Hubei Anhui           & 396  \\ 
8  &  Yunnan Guangxi Fujian       & 301  \\ 
9  &  Beijing Tianjin Hebei       & 429  \\ 
10 &  Others                      & 337  \\ 
\hline
\end{tabular}
\label{tab:region}
\end{table}

Both development set and test set are collected from our POI voice search production, Tencent Map.
The development set consists of 13,350 utterances collected from users across the whole China.
The test set contains 15,205 utterances collected from users of top-10 provinces with most traffic. 
The detailed data distribution in the development set and test set is presented in Table \ref{tab:data}.
\begin{table}[htp]
\caption{Number of utterances for each dialect region in the development set and the test set.}
\centering
\begin{tabular}{ccc}
\hline
\textbf{Regions} & \textbf{Development}  & \textbf{Test} \\
\hline
1 & 1744 & 3299 \\
2 & 1200 & 1300 \\
3 & 1986 & 2845 \\
4 & 1308 & 1910 \\
5 & 1201 & 1502 \\
6 & 801  & 1396 \\
7 & 1218 & 1519 \\ 
8 & 648  & 0    \\
9 & 1349 & 1434 \\
10 & 1895 & 0   \\
\hline
Total & 13350 & 15205 \\
\hline
\end{tabular}
\label{tab:data}
\end{table}

\subsection{Geo-AM}
\label{ssec:exp_am}

The baseline AM (A0) is trained with all corpora,
which is geo-location independent. We obtain a Geo-AM, A1, 
by attaching dialect-specific input feature to the baseline AM and
fine-tune it using corpora with dialect information. We also try to directly fine-tune A0 using the same corpora as A1, but achieve no improvement. Table~\ref{tab:am_dev} shows that our model benefits from the dialect-specific input feature, with an overall CER reduction of 4.3\%. Only slight improvement is found in some regions (e.g., Region 5, Region 6). We argue that division of dialect regions and distribution of training data should account for this.

For further improvement, a more superior Geo-AM A2 is obtained by introducing dialect-specific top layer to the model A1. The model A2 is initialized from A1. Dialect corpora are only used to train dialect-specific top layer while other parameters are frozen.
Results in Table~\ref{tab:am_dev} show that we can get more gains by introducing dialect-specific top layers. This indicates that Geo-vector is not powerful enough to encode dialect information.
Besides, 
dialect-specific top layers make it easy to improve the performance of a certain dialect, 
as we can train each top layer individually.

\begin{table}[htp]
\caption{CER (\%) of Geo-AMs on the development set.}
\centering
\begin{tabular}{cccccc}
\hline
\textbf{Region} & \textbf{A0} & \textbf{A1} & \textbf{A2} \\
\hline
1               & 4.93 & 4.73 & \textbf{4.63} \\
2               & 6.41 & 5.96 & \textbf{5.60} \\
3               & 5.23 & 4.93 & \textbf{4.80} \\
4               & 4.64 & 3.96 & \textbf{3.89} \\
5               & 4.98 & 4.91 & \textbf{4.86} \\
6               & 6.07 & 6.05 & \textbf{5.60} \\
7               & 6.21 & 6.03 & \textbf{5.71}\\
8               & 6.61 & 6.69 & \textbf{6.56} \\
9               & 4.01 & 3.79 & \textbf{3.73} \\
10              & 5.75 & \textbf{5.73} & 5.87 \\ 
\hline
Total           & 5.37 & 5.14 & \textbf{5.02} \\ 
\hline
\end{tabular}
\label{tab:am_dev}
\end{table}

To show A2's superiority,
we try to increase the amount of training data for dialect region 1,
from 522 hours to 892 hours.
Then we train another two Geo-AMs A1+ and A2+.
A1+ is fine-tuned from A0 and A2+ is fine-tuned from A1.
Results of Table~\ref{tab:am_dev_dialect1} show that both A1+ and A2+ achieve better performance on dialect region 1 compared to A1 and A2.
However, A1+ gets worse performance on several other dialect regions compared to A1 while A2+ maintains the performance on other dialect regions with the help of dialect-specific top layers. 
As we argued above, A2 gives us more flexibility which is important in real production service.

\begin{table}[htp]
\caption{CER (\%) of Geo-AMs on the development set after adding more data for dialect region 1.}
\centering
\begin{tabular}{ccccc}
\hline
\textbf{Region} & \textbf{A1} & \textbf{A1+} & \textbf{A2} & \textbf{A2+}  \\
\hline
1 & 4.73 & 4.59 & 4.63 & 4.45 \\ 
2 & 5.96 & 6.03 & 5.60 & 5.60 \\
3 & 4.93 & 5.15 & 4.80 & 4.80 \\
4 & 3.96 & 4.05 & 3.89 & 3.89 \\
5 & 4.91 & 4.90 & 4.86 & 4.86 \\
6 & 6.05 & 5.89 & 5.60 & 5.60 \\
7 & 6.03 & 6.14 & 5.71 & 5.71 \\
8 & 6.69 & 7.30 & 6.56 & 6.56 \\
9 & 3.79 & 3.76 & 3.73 & 3.73 \\
10 & 5.73 & 5.74 & 5.87 & 5.87 \\
\hline
Total & 5.14 & 5.20 & 5.02 & 4.99 \\
\hline
\end{tabular}
\label{tab:am_dev_dialect1}
\end{table}

To further verify the relationship between Geo-AM and multi-dialect problem,
we divide the development set into 4 subsets according to the level of accent and
provide results in Table~\ref{tab:am_dialect}.
It suggests that Geo-AM performs better on heavy-accent utterances.
\begin{table}[htp]
\caption{CER(\%) and CERR(\%) of Geo-AMs on datasets with different level of accent.}
\centering
\begin{tabular}{cccc}
\hline
\textbf{Level} & \textbf{A0} & \textbf{A2} & \textbf{CERR}  \\
\hline
Serious & 11.30 & 10.16 & 10.1 \\ 
Medium  & 9.46  & 8.67  & 8.4 \\ 
Slight  & 5.21  & 4.87  & 6.5 \\ 
None    & 3.72  & 3.54  & 4.8 \\ 
\hline
\end{tabular}
\label{tab:am_dialect}
\end{table}

\subsection{Geo-LMs}
\label{ssec:exp_lm}
Results in Section~\ref{ssec:exp_am} show that
the proposed Geo-AM can alleviate multi-dialect problem to some extent.
However,
it cannot deal with long-tailed and homophone POI names.
Therefore,
we adopt Geo-LMs in first-pass decoding as described in Section~\ref{sec:lm}.
Detailed results on the development set are presented in Table~\ref{tab:lm_dev}.
In order to evaluate whether Geo-LMs are effective in all provinces, 
Table~\ref{tab:lm_dev} provides the overall results as well as 
the results of top-5 provinces with the most traffic 
(Guangdong, Henan, Shandong, Jiangsu, Zhejiang) 
and tail-5 provinces with the least traffic 
(Gansu, Hainan, Ningxia, Xizang, Qinghai).
Results show that Geo-LMs can significantly improve the recognition accuracy of local POI names both in top provinces and tail provinces.
\begin{table}[htp]
\caption{CER (\%) on the development set of integrating Geo-LMs in first-pass decodeing (L1),
rescoring n-best lists of the first-pass decoding output without Geo-LMs (L2),
integrating Geo-LMs in n-best rescoring (L3).}
\centering
\begin{tabular}{lcccc}
\hline
\textbf{Province} & \textbf{A2} & \textbf{L1} & \textbf{L2} & \textbf{L3} \\
\hline
Guangdong & 4.88 & 4.56 & 4.35 & 4.42 \\
Henan     & 4.53 & 4.24 & 4.08 & 3.70 \\
Shandong  & 5.10 & 4.91 & 4.74 & 4.27 \\
Jiangsu   & 4.64 & 4.40 & 4.18 & 3.73 \\
Zhejiang  & 4.75 & 4.18 & 4.32 & 3.88 \\
\hline
Gansu     & 7.84 & 5.56 & 6.15 & 5.26 \\
Hainan    & 7.16 & 6.97 & 8.26 & 6.97 \\
Ningxia   & 7.19 & 6.47 & 6.29 & 6.12 \\
Xizang    & 5.88 & 5.88 & 5.88 & 4.24 \\
Qinghai   & 5.08 & 3.86 & 4.47 & 3.25 \\
\hline
Nationwide & 5.02 & 4.51 & 4.48 & 3.90 \\ 
\hline
\end{tabular}
\label{tab:lm_dev}
\end{table}

Rescoring n-best lists of the first-pass decoding output with a neural LM generally provides further improvements.
We use a QRNN model and a 5-gram character-level ngram model to rescoring the n-best lists 
like Eq.(\ref{eq:lm2}) but without Geo-LMs.
Results are shown in the forth column in Table~\ref{tab:lm_dev}.
Rescoring n-best lists reduces the CER on 4 top provinces and 1 tail province but
increases the CER on 1 top province and 3 tail provinces.
This is probably due to both the QRNN model and the 5-gram character-level ngram model do not utilize geographical information.
Therefore,
we also integrate a group of Geo-LMs in the process of n-best rescoring as Eq.(\ref{eq:lm2}).
Results are shown in the last column in Table~\ref{tab:lm_dev}.
Results show using Geo-LMs in second-pass decoding can further improve recognition accuracy of local POI names both in top provinces and tail provinces. 

Finally,
we evaluate the proposed Geo-AMs and Geo-LMs on the test set.
Results are consistent with those on the development set and details are shown in Table~\ref{tab:am_eval}.
The proposed Geo-AM and Geo-LMs totally achieve a 18.7\%  relative  CER reduction.
\begin{table}[htp]
\caption{CER (\%) on the test set of using Geo-AMs and Geo-LMs.}
\centering
\begin{tabular}{lcccccc}
\hline
\textbf{Province} & \textbf{A0} & \textbf{A1} & \textbf{A2} & \textbf{L1} & \textbf{L2}  & \textbf{L3}\\
\hline
Jiangsu   & 5.92 & 6.11 & 5.9  & 5.57 & 5.27 & 5.06  \\ 
Zhejiang  & 4.78 & 4.73 & 4.25 & 4.11 & 3.94 & 3.85  \\ 
Sichuan   & 3.69 & 3.44 & 3.38 & 3.06 & 3.2  & 3.18  \\ 
Shandong  & 3.75 & 3.74 & 3.6  & 3.54 & 3.05 & 3.06  \\ 
Henan     & 5.05 & 4.87 & 4.74 & 4.46 & 4.19 & 4.57  \\ 
Liaoning  & 4.33 & 4.02 & 3.94 & 3.44 & 3.36 & 3.0   \\ 
Guangdong & 5.9  & 5.54 & 5.5  & 5.22 & 4.86 & 4.68  \\ 
Shaanxi   & 4.98 & 4.87 & 4.55 & 4.45 & 4.0  & 3.88  \\ 
Anhui     & 4.92 & 4.82 & 4.56 & 4.73 & 4.26 & 4.16  \\ 
Hebei     & 3.93 & 3.7  & 3.65 & 3.43 & 3.37 & 3.11  \\ 
\hline
Total     & 4.70 & 4.58 & 4.38 & 4.17 & 3.99 & 3.82 \\ 
\hline
\end{tabular}
\label{tab:am_eval}
\end{table}


\section{CONCLUSION}
\label{sec:cons}
Speech recognition of local POI names is still a challenging task 
due to multi-dialect and massive long-tailed POI names.
This paper proposes a Geo-AM to deal with the multi-dialect problem 
by combining dialect-specific input feature and dialect-specific top layers.
In order to improve recognition accuracy of long-tailed POI names,
a group of Geo-LMs are integrated into the process of first-pass decoding and n-best rescoring.
Experiments show the proposed Geo-AM can indeed alleviate the accent problem and
achieve 6.5\%$\sim$10.1\% relative CER reduction on test sets of different accent levels.
The proposed Geo-AM and Geo-LMs totally achieve 18.7\%  relative  CER reduction on the POI voice search task of Tencent Map.
In addition,
introducing Geo-LMs into the process of n-best rescoring can achieve much better results.


\bibliographystyle{IEEEbib}
\bibliography{strings,refs}

\end{document}